\documentstyle[prl,aps,multicol,epsf]{revtex}        
\begin{document}
\tighten

\title{Ring-Pattern Dynamics in Smectic-C$^*$ and Smectic-C$_{\rm
A}^*$ Freely Suspended Liquid Crystal Films}
\author{D.~R.~Link, L.~Radzihovsky, G.~Natale, J.~E.~Maclennan, and
N.~A.~Clark} 
\address{Condensed Matter Laboratory, Department of
Physics, University of Colorado, Boulder, CO 80309 USA}
\author{M.~Walsh, S.~S.~Keast, and M.~E.~Neubert} 
\address{Liquid Crystal Institute, Kent State University, Kent, OH 44242 USA}

\date{\today} 
\maketitle

\begin{abstract}
  Ring patterns of concentric 2$\pi$-solitons in molecular
  orientation, form in freely suspended chiral smectic-C films in
  response to an in-plane rotating electric field.  We present
  measurements of the zero-field relaxation of ring patterns and of
  the driven dynamics of ring formation under conditions of
  synchronous winding, and a simple model which enables their
  quantitative description in low polarization DOBAMBC.  In smectic
  C$_{\rm A}^*$ TFMHPOBC we observe an odd-even layer number effect,
  with odd number layer films exhibiting order of magnitude {\em
    slower} relaxation rates than even layer films.  We show that this
  rate difference is due to much larger spontaneous polarization in
  odd number layer films.
\end {abstract}

\begin{multicols}{2}
\narrowtext

\begin{figure}[bth]
\centering
\setlength{\unitlength}{1mm}
\begin{picture}(150,50)(0,0)
\put(10,-15){\begin{picture}(150,0)(0,0)
\includegraphics{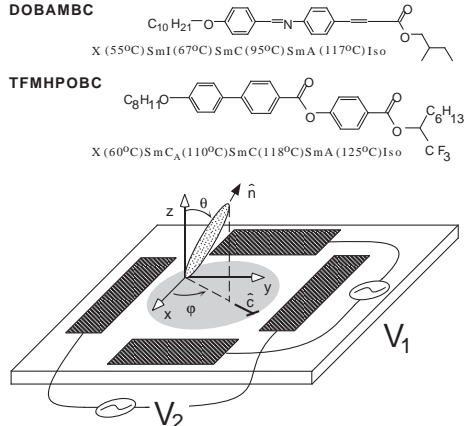}
\end{picture}}
\end{picture}
\caption{Chemical structures and phase diagrams for DOBAMBC and TFMHPOBC, and
  geometry for molecules in a tilted smectic phase.  The projection of
  the director {\bf n} onto the $x$-$y$ plane defines the {\bf
    c}-director ($\dashv$).  Freely suspended smectic liquid crystal
  films are drawn over a hole in a glass coverslip.  An in-plane
  rotating electric field is applied to the films by applying
  out-of-phase sinusoidal voltages, $V_1$ and $V_2$, to electrodes
  placed around the hole as shown.  The {\bf c}-director is visualized
  with depolarized reflected light microscopy.}
\label{geometry}
\end{figure}

Smectic liquid crystal phases are one-dimensional (1d) crystals
consisting of a periodic stack of two-dimensional (2d) liquid layers.
This rigidity allows them to be drawn into stable freely suspended
films, integer numbers of smectic layers thick.  In tilted smectic
phases, the rod-shaped molecules, their long axes on average pointing
along ${\bf n}$, strongly prefer to make an angle $\theta_0$ with the
layer normal ${\bf z}$, but have no energetic preference for the
global azimuthal orientation of {\bf n} about {\bf z}.  Consequently
in these tilted phases, this azimuthal low energy degree of freedom
can by described by a 2d unit director field, ${\bf c}(x,y)$, well
correlated between layers and pointing along the projection of ${\bf
  n}$ onto the smectic layer, as shown in Fig.~\ref{geometry}.
In chiral smectics, symmetry demands an in-plane polarization ${\bf
  P}(x,y)$ to be present, either along (longitudinal) or perpendicular
(transverse) to ${\bf c}(x,y)$, depending on the details of the tilted
smectic phase and the number of smectic layers in the
film\cite{Meyer,Link96,Link97}. Consequent coupling of the director
{\bf c}$(x,y)$ to an in-plane applied electric field has enabled
several key optical experiments on freely suspended films ranging from
light scattering\cite{LS} and microscopy\cite{PYC-80} to the
determination of ground state structure of novel liquid crystal
phases\cite{Link97}.

In these experiments, in order to prevent flow instabilities, the
direction of the in-plane aligning electric field must be reversed at
a frequency of 0.1 to 10 Hz. This process, still not well understood,
typically leads to the formation of beautiful concentric ring patterns
consisting of 2$\pi$-walls in ${\bf c}(x,y)$, which can be imaged via
Depolarized Reflected Light Microscopy (DRLM). A particular dark or
light region of a DRLM image has a {\bf c}-director ${\bf c}(x,y)=
[\cos(\varphi(r)),\sin(\varphi(r))]$ of constant azimuthal orientation
$\varphi$.  Thus, with $\varphi(r)$ pinned at the outer radius of the
film and monotonically increasing toward the center of the film, rings
appear. These distinctive ring patterns in ${\bf c}$ were first
studied quantitatively by Cladis et al.\cite{CladisNeedle}, who
generated them mechanically by rotating a needle inserted into the
center of a smectic-C (SmC) film.  Later it was reported that ring
patterns could be produced in ferroelectric films by applying a {\em
  rotating} (rather than bipolar) in-plane electric field ${\bf
  E}(t)$\cite{Hauck91}, with the first systematic experimental study
mapping out the ring-pattern formation phase diagram given in
Ref.\onlinecite{Koswig-93}.  Despite significant experimental
attention there has only been {\em qualitative} theoretical
understanding of ring-pattern formation and relaxation dynamics.  Here
we present a simple model, which enables quantitative understanding of
E-field wound ring-pattern dynamics and subsequent zero field ring
unwinding relaxation. We find excellent agreement with our experiments
on the low polariztion ($P\sim 3$~nC/cm$^2$) SmC*
DOBAMBC\cite{dobambc}.  However, our experiments on smectic-C$_{\rm
  A}^*$ TFMHPOBC\cite{tfmhpobc} reveal a novel odd-even smectic layer
effect, with order of magnitude slower relaxation times in the high
polarization odd layer number films and large ring distortion, driven
by bend/splay elastic anisotropy ($P\sim 75/N$~nC/cm$^2$, where $N$ is
an odd number of smectic layers). These latter observations lie beyond
our model and their quantitative description most certainly requires
understanding of screening-ion dynamics.

The free energy density of the 2d nematic orientation field {\bf
  c}$(x,y)$ of a freely suspended film in an applied field is given by
\begin{equation}
f(x,y)={K_S \over 2}(\nabla \cdot {\rm \bf c})^2 +{K_B \over2}({\rm
\bf \hat z}\cdot \nabla \times {\rm \bf c})^2-{\bf E}\cdot{\bf P}
 \label{eq:FE1}
\end{equation}
where the first two terms are the energies of splay and bend of {\bf
  c} with 2d elastic constants $K_S$ and $K_B$, respectively,
originating from the 3d Frank free energy for $\bf n$.  The last term
is the electric field director aligning energy, acting through the
polarization field density $\bf P$, which we take to be rigidly locked
to the $\bf c$ director, and for purposes of this developement,
orthogonal to {\bf c}.

Assuming simple relaxational dynamics for $\bf c$, with a viscous
damping coefficient $\gamma$, and taking $K_S\approx K_B\equiv K$ (but
see below), we obtain an equation of motion for $\varphi(x,y,t)$,
\begin{equation}
\gamma{\partial \varphi \over \partial t}=K\nabla^2\varphi - 
PE\sin (\varphi-\omega_e t)\;.
\label{eq-motion}
\end{equation}
In above, we have taken ${|\bf P|}\equiv P$ and ${|\bf E|}\equiv E$ to
be constants and the electric field rotating at frequency $\omega_e$
in the $x$-$y$ plane. Eq.~\ref{eq-motion} quite clearly ignores the
dipolar interaction and the rich 2d liquid hydrodynamics coupled to
the $\bf c$ director, which can in principle become important under
conditions of strong drive. Despite these shortcomings, as we will
show below, most of our experimental data on ring patterns dynamics in
the low $P$ liquid crystals is {\em quantitatively} described
by this model.  Its extension to treat observations in high
polarization materials will be a subject of a future
publication\cite{LinkRadzihovsky}.

A detailed analysis of Eq.~\ref{eq-motion} predicts a ring formation
phase diagram in the applied $E$-field strength and winding frequency
$\omega_e$ space, which is consistent with our experimental
observations. The dynamics is simplest in the regime of large $E\gg
E_c\equiv U_s/P$ and small $\omega_e \ll \omega_c\equiv P E/\gamma$,
in which the areal $\bf c$-director field alignment torque $P E$ is
much larger than {\em both} the pinning energy (torque) $U_s$ of $\bf
c$ at the outer boundary (radius $R$) of the film, and the frictional
torque $\gamma\omega_e$. In this regime the director field $\bf c$ is
{\em uniform}, synchronously following (at frequency
$\omega=\omega_e$) the rotating field $\bf E$, with a constant phase
lag $\delta=\sin^{-1}(\omega_e/\omega_c)$, set by the balance between
the $E$-field alignment and frictional torques. Clearly, for $\omega_e
> \omega_c$ this synchronous metastable solution, corresponding to the
areal alignment of {\bf P} along {\bf E} is unstable to a uniform
asynchronous dynamical regime, in which the director $\bf c$ uniformly
winds at a rate $\omega$ smaller than $\omega_e$ of the rotating $E$-
field. Away from the actual transition into this asynchronous regime,
the dynamics can be explicitly worked out perturbatively in
$\omega_c/\omega_e$ and we find that in addition to fast oscillatory
dynamics at harmonics of the ``washboard'' frequency $\omega_e$, the
spatially uniform phase $\varphi(t)$ advances {\em linearly} in time,
on average, with frequency
$\omega={1\over2}\omega_e(\omega_c/\omega_e)^2\ll \omega_e$, which,
interestingly, {\em decreases} with $\omega_e$.  Obviously, no rings
are produced in these two high $E$-field regimes.

Ring winding regimes lie in the range of low applied $E$ fields, such
that the pinning at the outer boundary (at $R$) is stronger than the
areal alignment torque $P E$. To analyze the synchronous,
$\omega_e<\omega_c$, ring winding dynamics, we take $\varphi({\bf
  r},t)\equiv\tilde{\varphi}({\bf r},t)+\omega_e t$, and look for
azimuthally symmetric traveling solution for $\tilde{\varphi}({\bf
  r},t)\equiv\vartheta(r-v t)$, which satisfies
\begin{equation}
\xi^2\ddot\vartheta +{v\over\omega_c}\dot\vartheta=
\sin\vartheta + {\omega_e\over\omega_c}\;,
\label{eqn:r}
\end{equation}
where ``dot'' indicates differentiation with respect to the argument
$r-vt$, $\xi\equiv\sqrt{K/PE}$, and we have, for now, neglected the
term $\xi^2\dot\vartheta/r$ that is subdominant for large rings.  The
boundary condition $\varphi(R,t)=\varphi_0$ translates into
$\vartheta=\varphi_0-\omega_e t$ and feeds in $2\pi$-solitons (winds
rings) at a rate $\omega_e$ from the outer boundary $r=R$ of the film
(see Fig.~2(a)).

Rings are traveling soliton solutions to the above equation, which can
be found by noting the isomorphicism of the Eq.~\ref{eqn:r} with the
Newtonian dynamics of a particle of mass $\xi^2$, friction coefficient
$v/\omega_c$ moving down a (unit strength) sinusoidal potential under
an external force $\omega_e/\omega_c$. First we note that, without
winding, a radial profile of an isolated ($\omega_e=0$) ring of radius
$r_0$ can be determined in closed form and is given by a well-known
soliton solution $\vartheta(r)=4\arctan\left[e^{(r-r_0)/\xi}\right]$.
It corresponds to the motion of a fictitious particle between two
maxima (at $0$ and $2\pi$) of the potential
$V(\vartheta)=1-\cos\vartheta$ {\em without} external force and
therefore a vanishing friction coefficient.  Consequently, without
winding, an isolated ring is stationary, $v=0$.  This stationary
$\omega_e=0$ solution can be easily extended to a concentric periodic
array of rings $d$ apart, by choosing the ``total energy'' $E$ and
therefore the initial ``kinetic energy'', $\xi^2(\dot\vartheta)^2/2$,
such that the effective particle can travel between maximas in
``time'' $d$. Although the solution can be expressed in terms of
special functions, the only features of it that are important to us is
its sigmoidal shape, the ring's radial width $\xi$ and
$\dot\vartheta|_{2\pi n}\approx 2\pi/d$.

For a finite synchronous winding rate, $\omega_e\neq 0$, the fictitious
particle is under an external constant force $\omega_e/\omega_c$,
leading to a tilted periodic potential. The mapping of the ring
winding problem onto particle dynamics makes it immediately clear that
in this case, a solution of a {\em periodic} array of self-similar
concentric rings is only possible if the fictitious particle moves in
the presence of a {\em unique} value of the frictional coefficient
$v/\omega_c$. This value is determined by the condition that the
``energy'' $2\pi\omega_e/\omega_c$, gained by the particle from
descending to a next local potential maximum is precisely the energy
$v/\omega_c\int_{\vartheta_0}^{\vartheta_0+2\pi}d\vartheta\dot\vartheta$
dissipated due to ``friction''. This condition predicts that, even in
the absence of tension, rings wound at rate $\omega_e$ {\em must} move
toward the center with velocity $v\approx(\pi/4)\omega_e\xi$, a result
that can also be clearly seen from noting that the $2\pi$-soliton
shifts by its width $\xi$ at frequency $\omega_e$, the rate of
rotation of the $\bf c$-director (see Fig.~2(a)).

\begin{figure}
\centering
\setlength{\unitlength}{1mm}
\begin{picture}(150,75)(0,0)
\put(-10,-10){\begin{picture}(150,0)(0,0)
\includegraphics{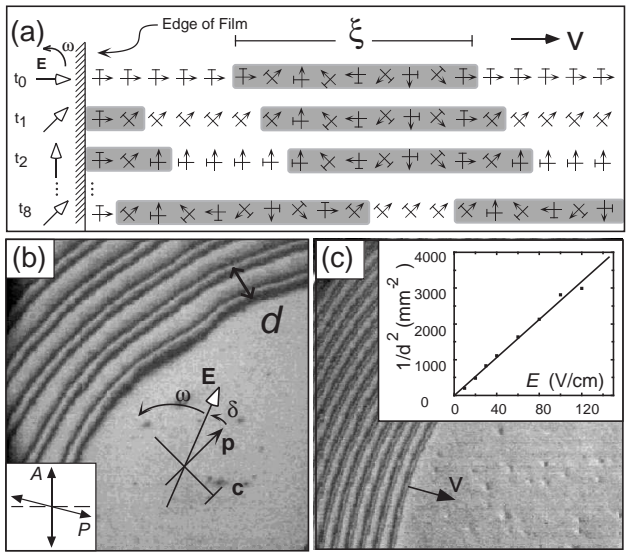}
\end{picture}}
\end{picture}
\caption{Synchronous-winding ring patterns. (a) At the edge of the film,
  boundary conditions prevent {\bf c} from rotating with the field
  resulting in the generation of a new 2$\pi$-soliton ring with every
  revolution of the field. A one dimensional 2$\pi$-wall (shaded region)
  in a rotating electric field {\bf E} moves with a constant velocity
  proportional to its width and rotation frequency of the field.  (b)
  Photomicrograph of the development of a ring pattern ($E=30$ V/cm).
  The separation between rings $d$ is independent of the frequency of
  rotation $\omega$ in the accessable frequency range (0.1 to 10 Hz).
  In (c) $E$ is increased by a factor of four ($E=120$ V/cm) reducing
  $d$ by a factor of two.  The inset in (c) shows the expected linear
  dependence of $1/d^2$ as a function of field strength.}
\label{fig:velocity}
\end{figure}

In the time $\tau_e=2\pi/\omega_e$ that it takes a new soliton ring to
be created at the outer film edge $R$, rings created before it, move
toward the center of the film a distance $d=v\tau_e=(\pi^2/2)\xi$,
predicting a steady state pattern of evenly spaced (by $d$), moving,
concentric rings, as illustrated in Fig.~\ref{fig:velocity}(b, c).  As
shown in the inset of Fig.~\ref{fig:velocity}(c), our experiments
indeed find $d^2\propto 1/E$ in agreement with the above theoretical
prediction for $d$ and $\xi$. We also, however, find that rings
spacing increases toward the center of the film, inconsistent with the
above $r$-independent prediction for $d$. It is easy to show that this
deviation is due to the increased importance (at small $r$) of the
ring line tension, contained in the $\xi^2\dot\vartheta/r$ term,
neglected in the Eq.~\ref{eqn:r}.  Line tension contributes an
additional $E$-independent velocity $\delta v_\tau={dr / dt}=-K /
(\gamma r) $ that must be superimposed on the velocity due to the
rotating field\cite{PYC-80}, and predicts a parabolic spacing of
solitons rings in the central region.

Contrary to the observations of Dascalu et al. \cite{Koswig-96}
experimentally we find both synchronous ($\omega < \omega_c$) and
asynchronous ($\omega > \omega_c$) winding of ring patterns. However,
in contrast to the well-defined soliton-like rings of width $\xi$
wound in the synchronous, $\omega_e<\omega_c$ regime discussed above,
in the asynchronous, $\omega_e>\omega_c$ regime, the rings are not
solitons and their width is roughly set by $R/n(t)$, decreasing as
their number $n(t)$ grows with frequency
$\omega={1\over2}\omega_e(\omega_c/\omega_e)^2$.  We have confirmed
experimentally our prediction of the linear $E$-field dependence of
the critical frequency $\omega_c$, separating these two ring winding
regimes.

We now turn our attention to ring unwinding dynamics at $E=0$ . We
find that, in contrast to previous claims in the
literature\cite{CladisNeedle} a general solution for the relaxation is
given by
\begin{equation}
\varphi(r,t)=\sum_n A_n J_0\left({a_{n} \over R}r \right)e^{-t/\tau_n}\;,
\label{eq:link}
\end{equation}
where $\tau_n=(\gamma/ K) ( R / a_{n} )^2$ is the time constant for
$n$th mode, $a_{n}$ are the zeros of the zeroth order Bessel function
$J_0(r)$, and $A_n$ are completely determined by the initial condition
$\varphi(r,0)$.  Since higher order ($n>1$) terms relax with a larger
time constant than lower order terms, $\varphi(r,t)$ rapidly takes on
the shape of the lowest $n=1$ term.

\begin{figure}
\centering
\setlength{\unitlength}{1mm}
\begin{picture}(150,60)(0,0)
\put(0,-15){\begin{picture}(150,0)(0,0)
\includegraphics{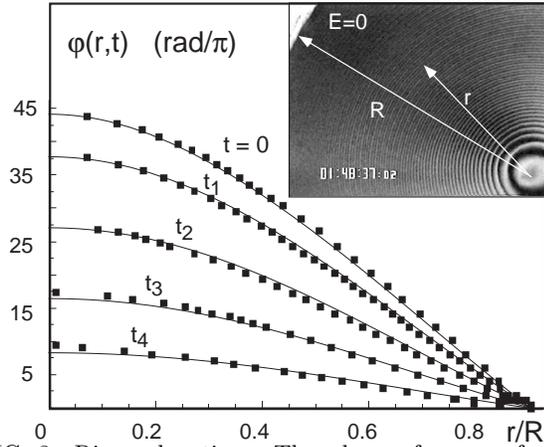}
\end{picture}}
\end{picture}
\caption{Ring relaxation.  The phase
of {\bf c} as a function of $r$ and $t$ is shown
(solid squares) for a five layer film of DOBAMBC during the relaxation
process.  The solid lines are fits to the first three terms of Eq.
(\ref{eq:link}) using
$A_1=134.9$,
$A_2=0.29$,
$A_3=1.15$ and
$\gamma/K = 1.99 \times 10^5$ s/cm$^2$ at times $t_1= 41.8$, $t_2=134.7$,
$t_3=275.1$ and
$t_4=467.4$ s.  The inset is a typical video image of a film or radius $R$ 
during
relaxation.}
\label{fig:DOBAMBC-relax}
\end{figure}

As a case study, we measured the relaxation of ring patterns in
ferroelectric SmC* DOBAMBC and antiferroelectric SmC$_{\rm A}^*$
TFMHPOBC.  Having determined $A_n$'s from the initial measured $E=0$
$\varphi(r,0)$ director profile, the subsequent evolution of
$\varphi(r,t)$ is completely specified by Eq.~\ref{eq:link}. As we
show in Fig.~\ref{fig:DOBAMBC-relax}, our theoretical prediction for
$\varphi(r,t)$ is in excellent agreement with the experimentally
determined evolution of the director profiles.  These one parameter
fits to experimental data allows us to quite accurately determine the
important ratio $\gamma/K$.  In a 5 layer film we find $\gamma/K=1.997
\times 10^5 {\rm s/cm^2}$, which, surprisingly, is an order of
magnitude larger than the previously reported value of $1.4 \times
10^4 {\rm s/cm^2}$\cite{PYC-80}, for reasons that are still unclear to
us.

\begin{figure}
\centering
\setlength{\unitlength}{1mm}
\begin{picture}(150,45)(0,0)
\put(-10,-25){\begin{picture}(150,0)(0,0)
\includegraphics{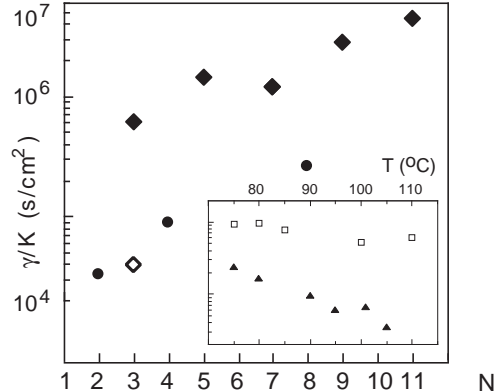}
\end{picture}}
\end{picture}
\caption{Layer number dependence of $\gamma/K$ in SmC$_{\rm A}^*$ TFMHPOBC. 
The effective ratio of $\gamma/K$ for films of different
number of layers shows a strong odd-even layer number dependence, with high
polarization, $N$-odd films (solid diamonds) having a much larger ratio than
the low polarization, $N$-even films (solid circles).  The ratio in an $N=3$
film (open diamond) of nearly racemic SmC$_{\rm A}^*$ TFMHPOBC is similar to
$N$-even films.  The temperature dependence of
$\gamma/K$ is shown in the inset for three layer (open squares) and two layer
(solid triangles) films.}
\label{TFMHPOBC}
\end{figure}
Such quantitative measurements in SmC$_{\rm A}^*$ TFMHPOBC revealed an
intriguing odd-even dependence of $\gamma/K$ on layer number $N$.  In
these experiments the relaxation at the center of the pattern,
$\varphi(0,t)$ was recorded and fit to Eq.~\ref{eq:link} and used to
extract $\gamma/K$. The results shown in Fig.~\ref{TFMHPOBC}, reveal
that $N$-odd films relax much more slowly than $N$-even films.  The
two main differences between $N$-odd and $N$-even films are that
$N$-odd films have a large net transverse polarization, while $N$-even
films have a significantly smaller net longitudinal polarization (in
the tilt plane)\cite{Link96}. To demonstrate that it is the difference
in the magnitude of the polarization between the odd and even layer
films that is responsible for this novel effect, we measured the
$\gamma/K$ ratio via our ring relaxation technique in three-layer
SmC$_{\rm A}^*$ films of almost-{\em racemic} TFMHPOBC (a small amount
of chiral TFMHPOBC was added to racemic TFMHPOBC so that $N$-odd films
would have a small net polarization).  These low polarization
odd-layer films display large {\bf c}-director fluctuations and ring
pattern relaxation rates comparable to $N=2$ and $N=4$ films, and have
an effective $\gamma/K$ an order of magnitude lower than the
enantiamerically pure material.  This strong dependence on the
magnitude of the spontaneous polarization indicates, that while
Eq.~\ref{eq:link} captures the essence of ring pattern relaxation, it
is unable to account for the difference in the relaxation rates
between small and large polarization materials.  We note that the
dynamics of low-polarization $N$-even films are described much better
by Eq.~\ref{eq:link} than those of high-polarization $N$-odd films.
The temperature dependence of $\gamma/K$, however, is similar in both
$N$-even and $N$-odd as is shown in the inset of Fig.~\ref{TFMHPOBC}.

This work by was supported by NSF grants DMR96-14061, DMR89-20147,
DMR-9809555, DMR-9625111, NASA grant NAG3-1846, and DARPA contract
MDA972-90C-0037. L.R. also acknowledges support by the A.P. Sloan and
David and Lucile Packard Foundations.


%
%
%
%
\end{multicols}
\end{document}